\documentstyle[12pt]{article}
\def\PRL #1 #2 #3 {Phys.~Rev.~Lett.~{\bf #1}, #2 (#3)}
\def\PRD #1 #2 #3 {Phys.~Rev.~D~{\bf #1}, #2 (#3)}
\def\PLB #1 #2 #3 {Phys.~Lett.~{\bf B#1}, #2 (#3)}
\def\NPB #1 #2 #3 {Nucl.~Phys.~{\bf B#1}, #2 (#3)}

\newcommand{\gtap}{{\raise.3ex\hbox{$>$\kern-.75em\lower1ex\hbox{$\sim$}}}}
\newcommand{\ltap}{{\raise.3ex\hbox{$<$\kern-.75em\lower1ex\hbox{$\sim$}}}}

\begin{document}
\renewcommand{\thefootnote}{\fnsymbol{footnote}}
\include{psfig}
\begin{titlepage}

\rightline{hep-ph/9806286}
\medskip
\rightline{June 1998}
\bigskip\bigskip
\begin{center}
{\Large \bf Unitarity and the scale of fermion mass generation} \\
\medskip
\bigskip\bigskip\bigskip
{\large{\bf S.~J\"ager}\footnote{\footnotesize Address after Nov.~1998: 
        Physik-Department, Technische Universit\"at M\"unchen, 
        James-Franck-Strasse, 85748 Garching, Germany.}\footnote{\footnotesize
        Email: sjaeger@physik.tu-muenchen.de}
and {\bf S.~Willenbrock}\footnote{\footnotesize Email: willen@uiuc.edu}} \\ 
\medskip 
Department of Physics \\
University of Illinois \\ 1110 West Green Street \\  Urbana, IL\ \ 61801 \\
\bigskip 
\end{center} 
\bigskip\bigskip\bigskip

\begin{abstract}
We study the upper bound on the scale of fermion mass generation in a 
two-Higgs-doublet model.  If the model is weakly-coupled, the scale 
of fermion mass generation is much less than the Appelquist-Chanowitz unitarity
bound.  However, if we allow some dimensionless Higgs self-couplings 
to become large, the Appelquist-Chanowitz unitarity bound can be saturated.
The unitarity bound on the scale of top-quark mass generation is about 3 TeV,
which may be within the reach of future colliders.
\end{abstract}
\end{titlepage}

\setcounter{footnote}{0}
\renewcommand{\thefootnote}{\arabic{footnote}}

\newpage

Many years ago, Appelquist and Chanowitz \cite{AC}
derived an upper bound on the scale of fermion mass generation for a 
fermion of mass $m_f$:\footnote{The strictest bound is obtained by considering
the spin-singlet, weak-isosinglet, color-singlet amplitude \cite{MVW}.}
\begin{equation}
\Lambda_f < \frac{8\pi v^2}{\sqrt{3N_c}m_f}
\label{f}
\end{equation}
where $v^2 \equiv (\sqrt 2 G_F)^{-1}$ and $N_c = 3$ for quarks 
and 1 for leptons.  This bound was obtained by
considering the scattering process $f\bar f \to V_LV_L$ [$V=W,Z$; $L$ denotes
longitudinal polarization (helicity zero)] in the absence of a Higgs boson.  
The resulting amplitude grows linearly with the center-of-mass energy, and
violates the condition of tree-level unitarity at the energy $\Lambda_f$ given
in Eq.~(\ref{f}).

If this bound is relevant, it has interesting implications for future 
high-energy colliders.  The numerical value of the bound for the top quark,
with a mass of $173.9 \pm 5.2$ GeV \cite{TOP}, is approximately 3 TeV. 
This energy could serve as a benchmark for the study of top-quark 
mass generation via $V_LV_L \to t\bar t$
at future hadron \cite{EMNN,DW,Y,LY},\footnote{At a hadron collider, one can 
also probe $t\bar t \to V_LV_L$, either at tree level through 
$gg \to t\bar tV_LV_L$ \cite{BPS} or at one loop through
$gg\to V_LV_L$ via a top-quark loop \cite{DKR,GV,BC,BBDKW}.} 
$e^+e^-$ \cite{Y,K,B,LY,LTY}, and $\mu^+\mu^-$ \cite{BBGH}
colliders, in much the same way that 
\begin{equation}
\Lambda_{EWSB}^2 = 8\pi v^2 \approx (1.2\; {\rm TeV})^2
\label{ewsb}
\end{equation}
(obtained from the consideration of $V_LV_L \to V_LV_L$ in the absence of a 
Higgs boson \cite{CG,MVW})
serves as the benchmark for electroweak symmetry breaking (EWSB).

The relevance of the bound in Eq.~(\ref{f}) was already questioned
in the original paper of Appelquist and Chanowitz.  In the standard Higgs
model, the Higgs boson is responsible for fermion mass generation, yet
its mass cannot exceed approximately 
800 GeV \cite{LW},\footnote{This upper bound is
obtained by demanding that the Higgs mass be less than the intrinsic cutoff
of the theory \cite{DN}.} well below the 
bound in Eq.~(\ref{f}) for all known fermions.  In extended Technicolor 
(ETC) models \cite{EL,DS}, the scale of fermion mass generation is the mass
of the ETC gauge boson
\begin{equation}
M_{ETC}\sim g_{ETC}\left(\frac{v^3}{m_f}\right)^{1/2} 
\label{etc}
\end{equation}
which scales with $m_f$ differently from
Eq.~(\ref{f}), and is much less than $\Lambda_f$ for all known fermions.  
Furthermore, the process $f\bar f \to V_LV_L$ ceases to
grow with energy at an energy of order $\Lambda_{EWSB}$
rather than $M_{ETC}$ \cite{AC}.  Several years later, Golden \cite{G}
revisited the issue of the relevance of the bound in models with a Higgs boson,
and found a bound which scales with $m_f$ like Eq.~(\ref{etc}) rather than 
Eq.~(\ref{f}).  Thus the relevance of the bound in Eq.~(\ref{f})
remains an open and important question.

In this paper we study the scale of fermion mass generation in the context of
a two-Higgs-doublet model.  
We imagine that the Higgs scalar responsible for
unitarizing $V_LV_L \to V_LV_L$ is ``light'',\footnote{A ``light'' Higgs boson
is one whose mass is less than about 800 GeV.} and we derive an upper bound
on the mass of the heavy Higgs scalar which is responsible 
for unitarizing $f\bar f\to V_LV_L$.  We find that if
all dimensionless Higgs self-couplings are kept fixed, 
the bound on the mass of the heavy Higgs scalar scales
with $m_f$ like Eq.~(\ref{etc}).  However, if some of the dimensionless
Higgs self-couplings
are allowed to grow with the mass of the heavy Higgs scalar, the bound on 
the mass of this Higgs scalar is instead given by the Appelquist-Chanowitz 
bound, Eq.~(\ref{f}). 

Before we begin our analysis, let us recall the meaning of unitarity bounds
in effective field theories \cite{AC,W}.  
In the absence of a Higgs boson, a perturbative calculation of the scattering
amplitude for $V_LV_L\to V_LV_L$ yields an expansion in powers of $E^2/v^2$.
This expansion becomes useless once the expansion parameter is of order unity.
One can use unitarity to estimate this energy \cite{CG,MVW}, which yields  
$E\sim \Lambda_{EWSB}$ (Eq.~(\ref{ewsb})).  The addition of a ``light'' 
Higgs boson to the theory
changes the expansion parameter to $m_h^2/v^2$ (for $E^2>>m_h^2$); 
one says that the Higgs boson
``unitarizes'' the theory.  The largest Higgs mass 
allowed is approximately 800 GeV \cite{LW}, which nearly saturates the
bound $\Lambda_{EWSB}$.

Now add a massive fermion of mass $m_f$ to the theory, again in the 
absence of a Higgs boson.  In processes involving
the massive fermion, one finds an additional expansion parameter, $m_fE/v^2$,  
along with the expansion parameter $E^2/v^2$.  Again, unitarity can be used
to estimate the energy at which this additional expansion parameter is of
order unity, and the result is $E\sim \Lambda_f$ (Eq.~(\ref{f})).  This
interpretation of the Appelquist-Chanowitz bound is supported
by the analysis of one-loop diagrams in section 2 of Golden \cite{G}.  

Whatever physics unitarizes
$V_LV_L\to V_LV_L$ need not necessarily unitarize $f\bar f\to V_LV_L$. 
In the standard Higgs model, the same Higgs boson is responsible for 
unitarizing both processes.  We therefore
consider a two-Higgs-doublet model, with only one Higgs field coupled to 
a given fermion.  This can be arranged by imposing a discrete symmetry 
$\Phi_1 \to -\Phi_1$ on one of the Higgs fields, such that only the Higgs
field $\Phi_2$ has Yukawa couplings to a given 
fermion \cite{HKS}.\footnote{Alternatively, one can couple a given fermion $f$
to the Higgs field $\Phi_1$ by letting $f_R \to - f_R$ under the discrete 
symmetry.}  The most general CP-symmetric scalar
potential, with the discrete symmetry $\Phi_1 \to -\Phi_1$ softly broken, 
is \cite{G1,HN,H}
\begin{eqnarray}
V(\Phi_1,\Phi_2) & = & m_{11}^2\Phi_1^{\dag}\Phi_1 
+ m_{22}^2\Phi_2^{\dag}\Phi_2
- m_{12}^2[\Phi_1^{\dag}\Phi_2 + \Phi_2^{\dag}\Phi_1] \nonumber \\
& + & \frac{1}{2}\lambda_1(\Phi_1^{\dag}\Phi_1)^2 
+ \frac{1}{2}\lambda_2(\Phi_2^{\dag}\Phi_2)^2
+ \lambda_3(\Phi_1^{\dag}\Phi_1)(\Phi_2^{\dag}\Phi_2)
+ \lambda_4(\Phi_1^{\dag}\Phi_2)(\Phi_2^{\dag}\Phi_1) \nonumber \\
& + & \frac{1}{2}\lambda_5[(\Phi_1^{\dag}\Phi_2)^2 + (\Phi_2^{\dag}\Phi_1)^2]
\end{eqnarray}
where all parameters are real.
The particle content of this model is two neutral Higgs scalars, 
$h^0$ and $H^0$; 
a neutral Higgs pseudoscalar, $A^0$; and a charged Higgs scalar, $H^{\pm}$.  
Only the neutral
Higgs scalars, $h^0$ and $H^0$, will be relevant for the ensuing discussion.

Our analysis of the model closely follows that of Haber \cite{H}.
The two Higgs fields acquire vacuum-expectation values $v_1$ and $v_2$,
where $v_1^2 + v_2^2 = v^2$, and we define 
$\tan\beta\equiv v_2/v_1$, as usual.  
The neutral-Higgs-scalar mass eigenstates are related to the weak 
eigenstates by
a rotation through the angle $\alpha$.  Their masses are given by
\begin{equation}
m_{H^0,h^0}^2 = \frac{1}{2}\left(m_S^2 
\pm \sqrt{m_S^4-4m_{A^0}^2m_L^2-4m_D^4} \right)
\label{mh}
\end{equation}
where
\begin{eqnarray}
m_{A^0}^2 & = & \frac{m_{12}^2}{s_\beta c_\beta}-v^2\lambda_5 \nonumber \\
m_L^2 & \equiv & v^2[\lambda_1c_\beta^4 + \lambda_2s_\beta^4
+2(\lambda_3+\lambda_4+\lambda_5)s_\beta^2 c_\beta^2] \nonumber \\
m_D^2 & \equiv & v^2[\lambda_5(\lambda_1c_\beta^4 + \lambda_2s_\beta^4) 
+ (\lambda_1\lambda_2 + \lambda_5^2 - (\lambda_3 + \lambda_4)^2)
s_\beta^2c_\beta^2]^{1/2} \nonumber \\
m_T^2 & \equiv & v^2(\lambda_1c_\beta^2 + \lambda_2s_\beta^2 + \lambda_5)
\nonumber \\
m_S^2 & \equiv & m_{A^0}^2 + m_T^2 
\label{ms}
\end{eqnarray}
and where we have adopted the short-hand notation $s_\beta \equiv \sin\beta$,
$c_\beta \equiv \cos\beta$.

The relevant Feynman rules for the Higgs scalars are listed in Fig.~1.
Since we want the light Higgs scalar $h^0$ to (almost completely) unitarize 
$V_LV_L\to V_LV_L$,
we want its coupling to vector bosons to be close to the standard-model
value.  Thus we are interested in the limit $\cos(\beta-\alpha)<<1$, 
where \cite{H}
\begin{equation}
\cos^2(\beta-\alpha) = \frac{m_L^2 - m_{h^0}^2}{m_{H^0}^2 - m_{h^0}^2}\;.
\label{cosba}
\end{equation}

\begin{figure}[t]
\centering
\setlength{\unitlength}{1in}
\begin{picture}(5,4)
\put(0,0) {\includegraphics{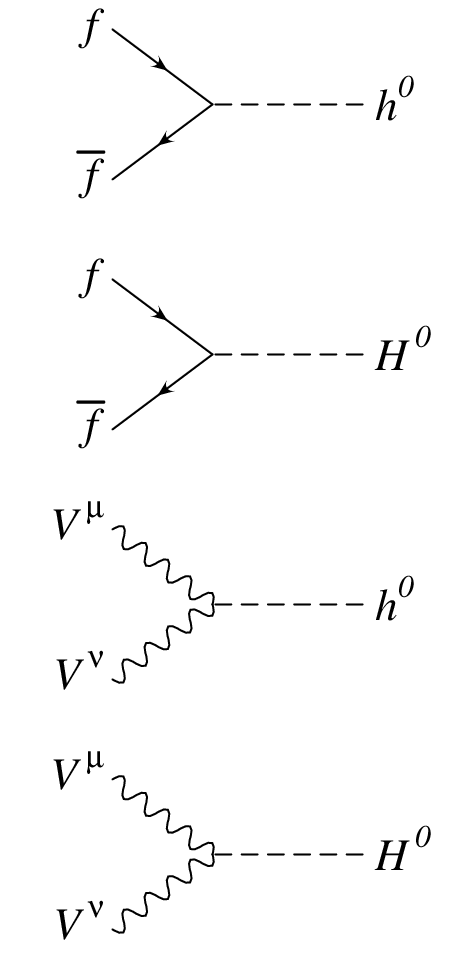}}
\put(1.8,3.45) { \parbox{0.1in}{ \[
   -i \frac {m_f} {v} \frac {\cos \alpha} {\sin \beta}
   = -i \frac{m_f}{v} [\sin(\beta-\alpha)+\cot \beta \cos(\beta-\alpha)]
   \] } }
\put(1.8,2.45) { \parbox{0.1in}{ \[
   -i \frac{m_f}{v}\frac {\sin \alpha} {\sin \beta}
   = -i \frac{m_f}{v} [\cos(\beta-\alpha)-\cot \beta \sin(\beta-\alpha)]
   \] } }
\put(1.8,1.45) { \parbox{0.1in}{ \[
   2 i \frac{M_V^2}{v} g^{\mu \nu} \sin (\beta-\alpha)
   \] } }
\put(1.8,0.45) { \parbox{0.1in}{ \[
   2 i \frac{M_V^2}{v} g^{\mu \nu} \cos (\beta-\alpha)
   \] }  }
\end{picture}
\caption{\small Couplings of the neutral Higgs scalars of a two-Higgs-doublet
model to fermion-antifermion and weak-vector-boson pairs.}
\end{figure}

We show in Fig.~2 the Feynman diagrams which, taken together, 
unitarize $f\bar f \to V_LV_L$.  The diagrams have a value equal to the
same diagram in the standard Higgs model, times the factor shown to the right
of each diagram.  Since we are interested in the limit 
$\cos(\beta-\alpha) <<1$, the only way that the diagram involving the 
heavy Higgs scalar
$H^0$ can contribute significantly to the unitarization of 
$f\bar f \to V_LV_L$ is if
\begin{equation}
\cot\beta \sin(\beta-\alpha) \cos(\beta-\alpha) \sim {\cal O}(1)
\end{equation}
or equivalently, since $\sin(\beta-\alpha) \approx 1$,
\begin{equation}
\cot\beta \sim 1/\cos(\beta-\alpha) >>1\;.
\label{beta}
\end{equation}
This corresponds to a strong Yukawa coupling of the heavy Higgs scalar to the
fermion (see Fig.~1), proportional to $(m_f/v)\cot\beta$.

\begin{figure}[t]
\centering
\setlength{\unitlength}{1in}
\begin{picture}(4.6,2)
\put(0,0) {\includegraphics{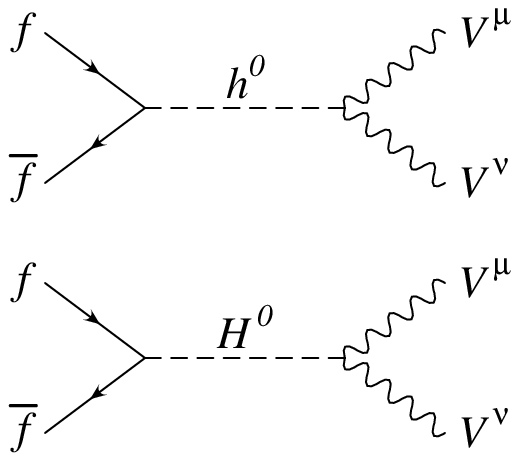}}
\put(2.3,1.45) { \parbox{0.1in}{ \[
   \sin^2(\beta-\alpha) + \cot \beta \sin(\beta-\alpha)\cos(\beta-\alpha)
   \] } }
\put(2.3,0.45) { \parbox{0.1in}{ \[
   \cos^2(\beta-\alpha) - \cot \beta \sin(\beta-\alpha)\cos(\beta-\alpha)
   \] }  }
\end{picture}
\caption{\small Diagrams which unitarize $f\bar f \to V_LV_L$ in a 
two-Higgs-doublet model.}
\end{figure}

To proceed further, we must introduce an additional piece of physics. 
Recall that the upper bound on the Higgs mass arises from the condition that
it be less than the cutoff of the theory \cite{DN}.  This condition
turns out to be equivalent to the condition that the (running) 
Higgs self-interaction
remain perturbative below the cutoff \cite{LW}.  Similarly, the Yukawa coupling
of a fermion must remain perturbative below the cutoff \cite{EG,S}.  
This puts an upper bound on the Yukawa coupling, which can be estimated
from the unitarity of $f\bar f\to f\bar f$ \cite{CHF,MVW}.  One finds
\begin{equation}
\frac{m_f}{v}\cot\beta < \left(\frac{4\pi}{N_c}\right)^{1/2}\;.
\label{yuk}
\end{equation}
Combining Eqs.~(\ref{beta}) and (\ref{yuk}) gives
\begin{equation}
\cos(\beta-\alpha) > \left(\frac{N_c}{4\pi}\right)^{1/2}\frac{m_f}{v}\;.
\label{cos0}
\end{equation}
This is a constraint on how small $\cos(\beta-\alpha)$ can be, and still have
the heavy Higgs scalar $H^0$ contribute significantly to the unitarization of
$f\bar f \to V_LV_L$.

To convert Eq.~(\ref{cos0}) 
into an upper bound on the mass of the heavy Higgs scalar $H^0$, we
must consider how $\cos(\beta-\alpha)$ scales with $m_{H^0}$.  If we hold
the dimensionless Higgs self-couplings $\lambda_i$ fixed, and increase 
the mass of the heavy Higgs scalar by increasing the parameter $m_{12}$,
Haber \cite{H} has shown that\footnote{This follows from Eq.~(\ref{cosba}),
due to a cancellation between the two terms in the numerator for large
$m_S^2$.}
\begin{equation}
\cos^2(\beta-\alpha) \sim \left(\frac{\lambda_iv^2}{m_{H^0}^2}\right)^2\;.
\label{cos1}
\end{equation}
Combining Eqs.~(\ref{cos0}) and (\ref{cos1}) yields
\begin{equation}
m_{H^0} < \left(\frac{4\pi}{N_c}\right)^{1/4}
\left(\frac{\lambda_iv^3}{m_f}\right)^{1/2}
\label{decouple}
\end{equation}
which scales with $m_f$ like Eq.~(\ref{etc}).  This is an example of
a model in which the scale of fermion mass generation scales with $m_f$ 
as in extended Technicolor, yet the process $f\bar f \to V_LV_L$ is unitarized
at the scale of fermion mass generation, $m_{H^0}$, rather than at 
$\Lambda_{EWSB}$, Eq.~(\ref{ewsb}).  The same scaling, but via a different
mechanism, is found in section 3 of Golden
\cite{G} from an analysis of higher-dimension operators; the model we are 
considering only involves operators of dimension four or less. 
For $\lambda_i \approx 4\pi$ (the largest value of the coupling 
such that it is perturbative), Eq.~(\ref{decouple}) yields an upper bound
of about 1.5 TeV on the scale of top-quark mass generation.

If we instead allow some of the $\lambda_i$ to increase with the mass of the 
heavy Higgs scalar, Haber \cite{H} has shown that we can have 
\begin{equation}
\cos^2(\beta-\alpha) \sim \frac{\lambda_iv^2}{m_{H^0}^2}
\label{cos2}
\end{equation}
instead of Eq.~(\ref{cos1}), while still keeping $m_{h^0}$ small.
This can be achieved by making $m_T^2 \sim m_S^2$ 
(see Eq.~(\ref{ms})),\footnote{This disrupts the
cancellation mentioned in the previous footnote, such that Eq.~(\ref{cosba})
results in Eq.~(\ref{cos2}).} 
while keeping $m_{A^0}^2m_L^2 + m_D^4 << m_S^2$ (see Eq.~(\ref{mh})).
A simple way to arrange this is to take 
$\lambda_5 \sim m_{H^0}^2/v^2$,
while keeping all other $\lambda_i$ fixed.\footnote{This yields 
$s_\beta^2 \sim v^2/m_{H^0}^2$, as follows from Eqs.~(\ref{beta}) 
and (\ref{cos2}), which is necessary to keep $m_L^2$ and $m_D^2$ small.}  
Combining Eqs.~(\ref{cos0})
and (\ref{cos2}) yields
\begin{equation}
m_{H^0} < \left(\frac{4\pi}{N_c}\right)^{1/2}\frac{\sqrt \lambda_i v^2}{m_f}
\end{equation}
which scales with $m_f$ like the Appelquist-Chanowitz bound, Eq.~(\ref{f}).
For $\lambda_i \approx 4\pi$ (the largest value of the coupling such that 
it is perturbative), the mass of the heavy Higgs scalar saturates
this bound.  This is the first example of such a model of which we are aware.
However, since {\em all} the dimensionless Higgs self-couplings 
(including $\lambda_5$) may be restricted to be perturbative, it is 
not clear how to interpret this model.

Since $\cos(\beta-\alpha)$ has a lower bound given by Eq.~(\ref{cos0}), 
we should check that the amplitude for $V_LV_L \to V_LV_L$ is 
sufficiently unitarized by the light Higgs scalar $h^0$ that it remains 
within the unitarity bound up to the mass of the heavy Higgs scalar $H^0$.
The $s$-wave, isosinglet amplitude is given by \cite{CG}
\begin{equation}
a_0^0 = \frac{s}{16\pi v^2}
\left[1-\frac{s}{s-m_{h^0}^2}\sin^2(\beta-\alpha)\right]
\to \frac{s}{16\pi v^2}\cos^2(\beta-\alpha)
\end{equation}
where the last expression is valid for $s>>m_{h^0}^2$.  Using 
Eq.~(\ref{cos0}), this implies
\begin{equation}
a_0^0 > \frac{s}{16\pi v^2}\frac{N_c}{4\pi}\frac{m_f^2}{v^2}
\end{equation}
which respects the unitarity condition $|{\rm Re}\;a_0^0| < 1/2$ up to the 
energy
\begin{equation}
\sqrt s < \frac{4\pi \sqrt 2 v^2}{\sqrt{N_c}m_f}\;.
\end{equation}
This is essentially the Appelquist-Chanowitz bound, Eq.~(\ref{f}), so we are
assured that the heavy Higgs scalar which unitarizes $f\bar f \to V_LV_L$
is also light enough to complete the unitarization of $V_LV_L \to V_LV_L$.

We have shown that in a weakly-coupled two-Higgs-doublet model, the scale of 
fermion mass generation is bounded by Eq.~(\ref{decouple}) rather than by
the higher Appelquist-Chanowitz bound, Eq.~(\ref{f}).  For the top quark,
this bound is about 1.5 TeV, which makes it an interesting benchmark for 
future colliders.  However, if we allow some of the dimensionless Higgs 
self-couplings 
to become large, then we are able to saturate the Appelquist-Chanowitz bound.
While this may not be physically realizable, it suggests that the 
Appelquist-Chanowitz bound may be relevant for strongly-coupled models
of fermion mass generation.  This bound is about 3 TeV for the top quark,
which is also an interesting benchmark for future colliders.

\section*{Acknowledgments}

\indent\indent We are grateful for conversations with T.~Han 
and J.~Shigemitsu. 
S.~W.~was supported in part by the U.~S.~Department 
of Energy under contract No.~DOE~DE-FG02-91ER40677.

\end{document}